# Performance-based ice engineering framework: a data-driven multi-scale approach

Reda Snaiki [a, *]

[a] Department of Construction Engineering, École de Technologie Supérieure, Université du Québec, Montréal, Quebec, Canada

[*] *Corresponding author email*: reda.snaiki@etsmtl.ca

**Abstract**: Ice storms are one of the most devastating natural hazards which have the potential to inflict significant damage to the built environment. The multi-hazard nature of ice events complicates the analysis of their induced risk due to their inherent nonlinear interactions. In addition, the concurrent and interacting hazards are often responsible for several aerodynamical/dynamical instabilities such as the galloping mechanism. Moreover, the existing risk approaches are usually not suited for large-scale risk evaluation over extended geographical regions due to the involved high-computational costs. Therefore, in this study, a novel data-driven multi-scale performance-based ice engineering (PBIE) framework is developed to support the design of new structures subjected to ice storms or the rehabilitation of existing ones. In addition, the proposed PBIE is capable of rapidly estimating the real-time risk over an extended region due to an ice event. Specifically, it leverages the superior capabilities of state-of-the-art data-driven techniques (e.g., machine learning) to efficiently generate the corresponding risk maps and identify the high-risk areas. The proposed PBIE framework is applied to a simplified example in which the galloping-induced risk on iced conductors, in terms of the galloping amplitude, is evaluated for both local and regional scales. The resulting PBIE framework can be readily applied for design or retrofitting purposes or integrated within an emergency response management system to inform preventive actions that can mitigate the ice storm-induced damages and save lives.

## 1. Introduction

Ice storms are severe weather events which can cause widespread and significant damage to several structures and systems including power line networks, buildings, bridges, wind turbines and transportation infrastructures (Fikke et al., 2008). For instance, ice accretion on overhead transmission line systems can result in galloping and eventually power outage (due to the possible collapse of these systems) (Lu and Chakrabarti, 2023). Similarly, stay cables might experience extreme galloping due to the joint wind and ice accretion effects which could eventually compromise the safety of cable-stayed bridges (Sharma et al., 2022). Ice accretion on the wind turbine blades can also result in power loss and possibly significant structural vibrations due to the changing aerodynamic characteristics. There have been several notable ice storm events over the past few decades which have caused significant damage to life and property. For example, the North American ice storm of 1998 has affected large parts of eastern Canada and the northeastern United States in January of 1998. This ice storm was responsible for extensive damage to power lines, buildings and other structures resulting in an estimated $4 billion US of damage, in Canada and the United States, with more than 4 million customers without electricity and 47 fatalities (Henson et al., 2011; Mittermeier et al., 2022). The December 2013 North American storm

complex was also a severe ice storm which struck the northeastern United States causing widespread power outages and massive damage due to fallen trees. Over a million residents were left without power due to the ice storm which has killed several people and resulted in almost $200 million of damage (Armenakis and Nirupama, 2014). With climate change, global temperatures are expected to rise which will affect the precipitation patterns and eventually, the frequency, intensity and duration of ice storms (Le Roux et al., 2021). Therefore, it is important to develop novel approaches which will ensure effective design and maintenance of new and existing structures in regions affected by ice storms or assist in the rapid estimation of high-risk regions for large geographical areas.

Performance-based design (PBD) is a state-of-art approach which was originally developed for earthquake engineering applications (Porter, 2003). It is a design methodology that aims to achieve predefined performance objectives, rather than meeting minimum standards and requirements or only optimizing/balancing safety and costs (Ciampoli et al., 2011). By prioritizing performance objectives, PBD can result in a structure that is safer, more resilient, and more sustainable while providing more flexibility in the design process. The PBD requires a comprehensive understanding of the potential hazards, structural systems and expected performance goals to accurately assess and reduce the hazard-induced risk and ensure that the structures and systems can perform as intended while meeting the performance objectives (Yu et al., 2020, Favier et al., 2022). Performance-based design has also been extended to other fields, giving rise to new PBD applications such as performance-based wind engineering (PBWE) (Ciampoli et al., 2011, Ouyang and Spence, 2021), performance-based coastal engineering (PBCE) (González-Dueñas and Padgett, 2021, Do et al., 2016), performance-based fire engineering (PBFE) (Rini and Lamont, 2008), performance-based tsunami engineering (PBTE) (Attary et al., 2021), and performance-based hurricane engineering (PBHE) (Barbato et al., 2013), among others. These frameworks are all based on the theorem of total probability in which the risk assessment is disaggregated into several modules (e.g., hazard and structural analysis). However, potential applications of the PBD to ice engineering have not yet been developed. In addition, existing PBD frameworks are usually developed for specific structures and are not suited for large-scale risk evaluation over extended geographical regions due to the involved high-computational cost.

In this study, a novel and flexible multi-scale performance-based ice engineering (PBIE) framework will be developed that can effectively estimate the real-time induced risk and support the design of structural systems. Specifically, the proposed PBIE framework has the ability to be used at both the structural (local) scale and the regional (large) scale. An overview of the proposed methodology will be first presented, followed by a description of the performance objectives & uncertainty assessment. The comprehensive PBIE will be then outlined in detail by covering each module of the framework. The proposed procedure is then applied for a case study in which the vertical galloping-induced risk on iced conductors will be evaluated. Specifically, the risk will be expressed in terms of the galloping amplitude. Machine learning techniques will be integrated within the regional scale to rapidly estimate the galloping induced risk.

## 2. Overview of the Proposed Multi-Scale PBIE

This study introduces a novel Performance-Based Ice Engineering (PBIE) framework designed to efficiently assess real-time risk from ice events and aid in the design of civil engineering structures. The PBIE framework is suitable for both structural (local) and regional (large) scales. It can be used to guide the design of new structures, assess the need for retrofitting existing ones, and even predict ice storm risks as they unfold in real-time by leveraging advanced data-driven techniques. Further details will be elaborated in the upcoming sections.

## 2.1 PBIE for the local scale

To estimate the ice storm-induced risk on a given structure (e.g., transmission tower-line systems and wind turbines), the structural risk is expressed in terms of the exceedance probability for a preselected decision variable ($DV$) corresponding to a predefined target performance (Ciampoli et al., 2011) as expressed in Eq. (1):

$$G(DV) = \int \int \int \int \int G(DV|DM).f(DM|EDP).f(EDP|IM,IP,SP).f(IP|IM,SP) \\ .f(IM).f(SP).dDM.dEDP.dIP.dSP.dIM \quad (1)$$

where $G(.)$ = complementary cumulative distribution function; $G(.|.)$ = conditional complementary cumulative distribution function; $f(.)$ = probability density function (PDF); $f(.|.)$ = conditional PDF; $DM$ = damage measure; $EDP$ = engineering demand parameter; $IM$ = intensity measure; $IP$ = interaction parameter; and $SP$ = structural parameter. To evaluate the exceedance probabilities using Eq. (1), the risk assessment can be disaggregated into the following steps which are depicted in Fig. 1: 1. hazard analysis, 2. structural characterization, 3. interaction analysis, 4. structural analysis, 5. damage analysis, 6. loss analysis, and 7. decision-making.

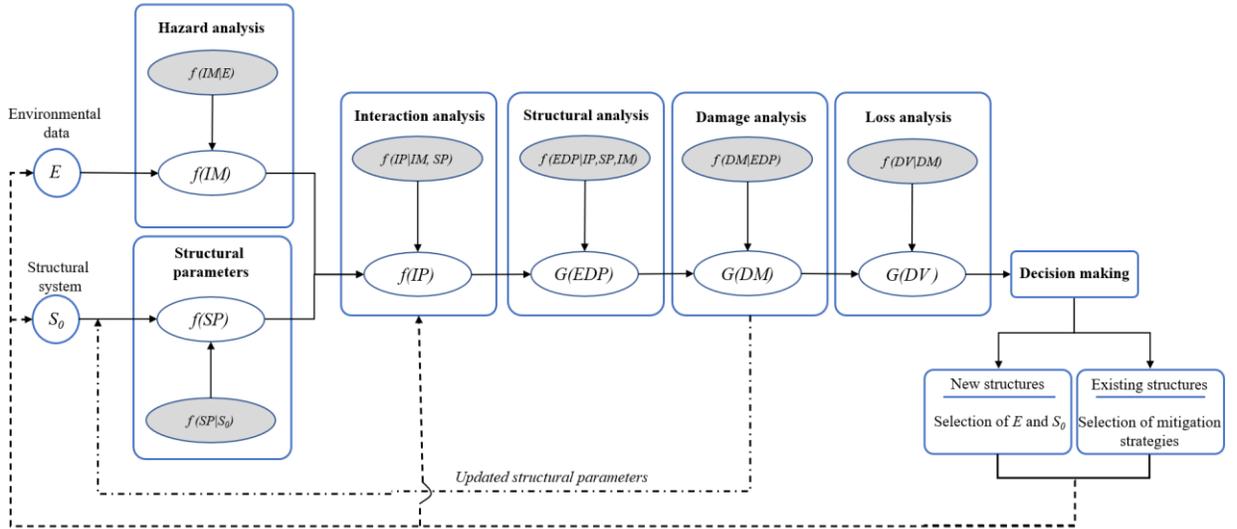

**Fig. 1**. Performance-based ice engineering framework

It is interesting to highlight that the proposed PBIE framework, as illustrated in Fig. 1, is applicable for both the design of new structures and for mitigating (also retrofitting) the effects of extreme events on the existing structures. For new structures, the decision-making step assists in identifying optimal candidate locations for the structure which should be characterized by reduced intensity measures. In addition, it will help in optimizing the structural parameters (e.g., geometric and mechanical properties) leading to acceptable structural response while minimizing the damage and loss (Li et al., 2021). On the other hand, for existing structures, the decision-making step will identify suitable mitigation strategies to minimize the risk by either changing the structural properties (e.g., added damping) or even by altering the environmental conditions (e.g., air injection in strategic locations to change the wind field). In addition, it can change the aerodynamics through several techniques (e.g., de-icing to change the accreted ice profile). It should also be noted that the framework can accommodate the change of structural properties due to the exceedance of predefined damage levels. The proposed PBIE framework is also flexible such that it is possible to account for time-dependent parameters, including those related to the environmental parameters (e.g., climate change effects) and the structural properties (e.g., aging effects).

## 2.2 PBIE for the regional scale

Since the ice events usually impact large geographical areas, the estimation of their induced risk becomes intractable due to the involved high-computational cost related to the employed models which simulate each module of the PBIE framework (Fig. 1). In addition, solving Eq. (1) to estimate the risk values is usually based on the computationally expensive Monte Carlo simulation which integrates the risk uncertainties in each risk module. Therefore, an alternative approach is proposed here for risk assessment on the regional scale based on data-driven techniques (e.g., artificial neural networks) to alleviate the high computational cost of the physics-based models and numerical techniques. Complementary techniques can also be included in this framework, such as the dimensionality reduction methodologies (e.g., principal component analysis and autoencoders) to reduce the high dimension of the predicted quantities which yields a low-dimensional latent space where the simulations can be carried out before projecting them back to the original space. Once the potential high-risk areas are identified using the regional scale-based PBIE, then the local scale-based PBIE can be used to accurately estimate the ice storm-induced risk by employing advanced models for each risk module. Therefore, the regional scale-based PBIE assists in the rapid identification of the vulnerable structures and provides a real-time risk analysis which can be embedded within an emergency response management system (Haimelin et al., 2017). The schematic implementation of the regional scale-based PBIE is depicted in Fig. 2.

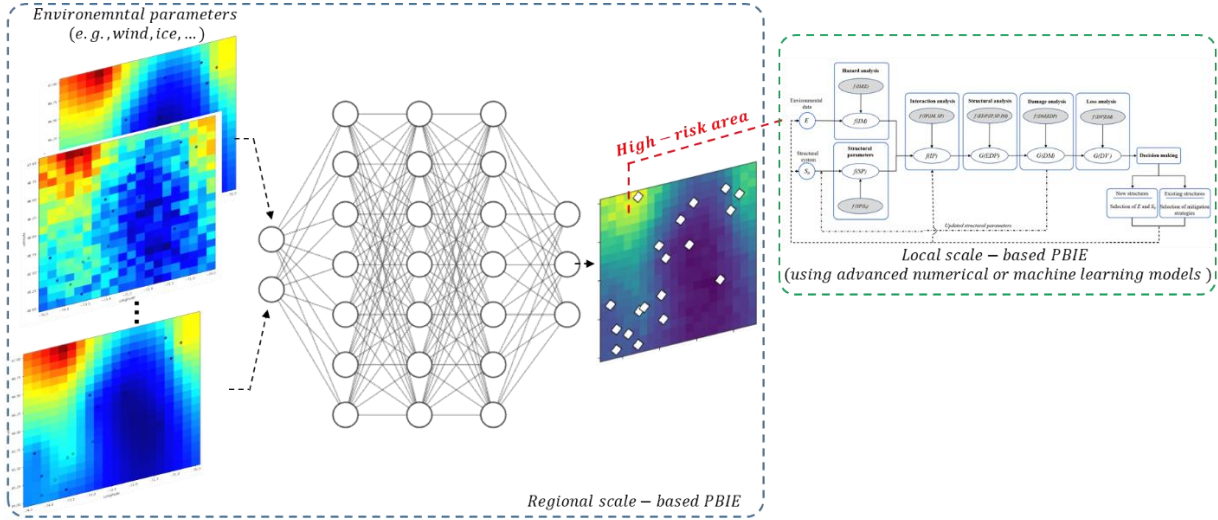

**Fig. 2**. Schematic implementation of the data-driven regional scale-based PBIE

## 3. Performance Objectives & Uncertainties Assessment

### 3.1 Performance objectives

Accurately defining the performance objectives of a particular structural system (e.g., transmission tower-line systems, wind turbines and cable-stayed bridges) is crucial for the performance design approach since the risk values are obtained by verifying if the performance objectives have been achieved or not. They are usually quantified in terms of a permissible level of damage (also denoted as performance level) of the structural system given a hazard event. Depending on several factors including the studied structure, hazard type, consideration of structural or non-structural elements, construction materials and the type of functionality itself, the definition of the performance expectation changes considerably (Fema, 2003). Traditionally, the performance objectives have been classified into two categories, namely low- and high-performance levels where the former is related to the structural safety and integrity (e.g., partial or total structure collapse) and the latter is related to the serviceability and comfort (e.g., vibration limitation) (Ciampoli et al., 2011). However, new categories have been recently suggested to complement the definition of the performance expectations. For instance, an additional performance objective related to the evacuation prior to severe ice storm events (e.g., for the case of cable-stayed bridges) might be accounted for. The performance objectives can also be expressed in terms of the operational status (e.g., wind turbines and transmission tower-line systems) and the sustained losses (Barbato et al., 2013). For example, three performance objectives can be defined within the PBIE framework for the case of transmission tower-line systems subjected to ice storms. The first one is the high-performance level which characterizes the fully operational status of the system where no damage to the structural or non-structural elements has occurred and the system continues working without interruption. The second category is the intermediate performance level which could also be divided into sub-levels in which the system is either fully or partially operational due to incurred damages to non-structural elements. The last category is the low performance level in which the transmission tower-line system is not operational due to the structural damage (could be extensive) which jeopardizes the structural integrity. It should be noted that each category could be further divided into sub-categories with detailed description on the damage level depending on several conditions (e.g., structure type and hazard).

### 3.2 Uncertainties assessment

The PBIE is based on several stochastic random variables which characterize each module of the framework (Fig. 1). Therefore, identifying and expressing those uncertainties is important to accurately estimate the risk values. In general, the uncertainties can be classified into two major categories, namely aleatoric and epistemic uncertainties (Der Kiureghian and Ditlevsen, 2009). While the aleatoric uncertainties are practically irreducible due to the randomness and unpredictability of the system, the epistemic uncertainties can be reduced using several approaches, including the improved knowledge, advanced modeling and extensive statistical analysis. The uncertainty sources can also be classified according to three main zones as depicted in Fig. 3:

1. *The environment zone*: In this zone, the hazard intensities are characterized without the consideration of the structure interference effects. However, for multi-hazard events (e.g., ice storms) an interaction between the hazards themselves could occur (e.g., momentum exchange between air and ice or raindrops) which might complicate the assessment of the involved uncertainties. The uncertain parameters in this zone are grouped in the intensity measure vector *IM*. Examples of *IM* uncertainties are the errors associated with field measurements (e.g., anemometer and satellite) or the hazard modeling (e.g., empirical, analytical or numerical models).

2. *The exchange zone*: This is the region surrounding the structural system in which the hazard and the structure are highly correlated. The interference effects due to the presence of nearby structures are also important within this zone. The uncertain parameters in this zone are grouped in the interaction parameter vector *IP*. Examples of *IP* uncertainties include the aerodynamic and aeroelastic characteristics (e.g., drag and lift coefficients; aerodynamic and aeroelastic derivatives). For instance, in the case of iced overhead lines, the new shape of the conductor due to the accumulated ice will affect the wind-induced aerodynamic load which will also depend on the angle of attack between the direction of the flow and the structure.

3. *The structural zone*: The uncertain parameters related to the structural system are grouped in the vector of structural parameters *SP*. Examples of *SP* uncertainties include the geometrical and mechanical properties of the structural system (e.g., stiffness, size, shape and orientation) (Koh and See, 1994). In addition, the non-environmental actions which can alter the structural system should be accounted for along with their uncertainties through the *SP* vector.

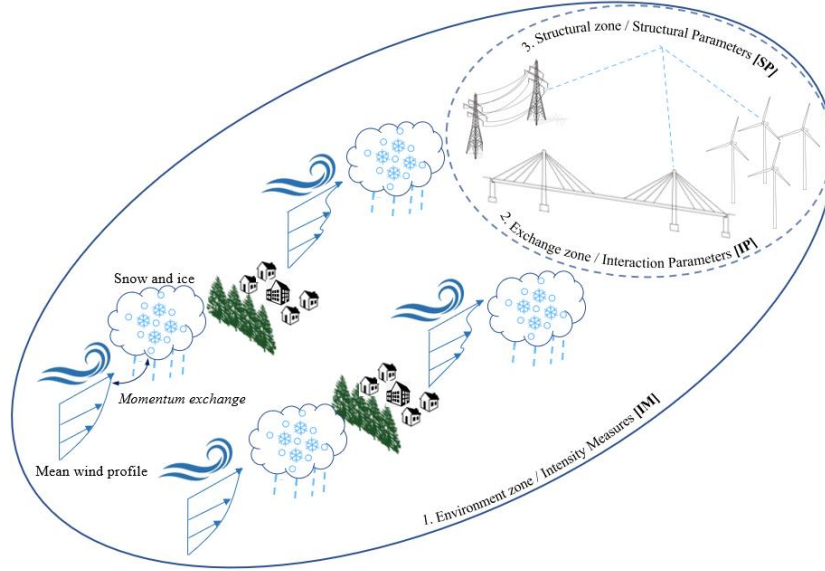

**Fig. 3.** Schematic of the source of uncertainties in the PBIE grouped by three main zones (Note: although only wind and ice hazards have been plotted in the figure, other hazards can be included [e.g., debris])

## 4. Detailed PBIE Procedure

In this section, a detailed description of the PBIE modules will be first covered including the hazard analysis, structural characterization, interaction analysis, structural analysis, damage analysis and loss analysis. Then, the practical implementation of the regional risk will be discussed.

### 4.1 Hazard analysis

In this step, the intensity measure vector *IM* is identified and characterized probabilistically to account for the involved uncertainties. Since ice events are usually multi-hazards in nature, the vector of intensity measures is expressed as:

$$IM = [\{IM_1\}, \{IM_2\}, \dots, \{IM_i\}, \dots, \{IM_n\}] \tag{2}$$

where $n$ = number of intensity measures involved in the ice storm (e.g., wind and ice). Each intensity measure $\{IM_i\}$ usually depends on the fundamental storm parameters '$\beta_i$' which can be expressed as:

$$\{IM_i\} = [\beta_i^1, \beta_i^2, \ldots, \beta_i^m]^T \tag{3}$$

where $m$ = number of storm parameters which define the intensity measure. For instance, the wind speed in extratropical cyclones can be determined using five storm parameters (Snaiki and Wu, 2022): central pressure deficit ($\Delta p$), storm transitional speed ($c$), approach angle ($\alpha$), surface roughness ($z_0$), and the latitude of the storm center ($\varphi$). Hence, the intensity measure can be expressed as $IM = [\Delta p, c, \alpha, z_0, \varphi]$. Once the intensity measures have been identified, their joint probability density function (PDF), $f(IM)$, which is required by the PBIE framework (Eq. 1), need to be determined. The joint PDF, $f(IM)$, can be estimated using several techniques such as copulas, Gaussian kernel density functions, and the generalized polynomial chaos. To generate the marginal probabilities of the intensity measures along with the joint PDF, $f(IM)$, sufficient data are required. Several techniques can be used in this regard to define the ice storm-related intensity measures (e.g., wind and ice):

1. *Direct statistical approach*: here, the historical data for the intensity measures are directly retrieved from existing databases at a specific site. The retrieved data are then used to fit selected probability distributions.
2. *Indirect statistical approach*: here, the historical data of the fundamental storm parameters '$\beta_i$' of Eq. (3) which define specific intensity measures are instead used and statistically represented by selected PDFs. Then, a Monte Carlo simulation is applied to sample and propagate their statistics in a given mathematical formulation which relates the storm parameters to the intensity measures.
3. *Synthetic track approach*: here, advanced statistical approaches are used to generate a large stochastic database of synthetic ice storms from genesis until dissipation. This technique enables the accurate simulation of extreme and rare events which are usually hard to capture and affect the tails of the probability distribution. The tracks are then combined with mathematical representation of the hazard intensity to generate the hazard probabilities.

It should be noted that, depending on the way by which the different hazards interact, the hazard analysis will be different. This point has been addressed by (Unnikrishnan and Barbato, 2017) where they classified the hazards into independent hazards, interacting hazards, and hazard chains. All these categories can be readily implemented within the proposed PBIE with proper identification of the joint PDF. During an ice storm event, both wind and ice hazards can cause enormous damage to civil infrastructures (e.g., transmission tower-line systems). Other hazards can be important depending on the selected structure and location. For instance, offshore wind turbines can be subjected, during an ice storm, to the effects of wind, ice, storm surge, waves, and currents. Windborne debris (also waterborne debris or debris avalanche) can also affect specific structures (e.g., buildings). In what follows, a brief description of both wind and ice hazards will be presented.

*Wind Hazard*: The wind speed can be decomposed into a mean and a turbulent component. The variation of the mean wind speed with height can be well represented using the logarithmic or power law. However, during an ice storm the mean wind speed profile does not necessarily follow the standard logarithmic law. In fact, due to several dynamical and thermodynamical processes, the vertical mean wind profile can present supergradient regions where the mean wind speed within the boundary layer can be higher than the gradient wind. Therefore, advanced numerical or data-driven models should be used instead to simulate the wind profile when the historical data is not enough to derive statistical estimations. On the other hand, the turbulent wind is estimated using statistical approaches. For the single-point statistics, the turbulent wind can be generated provided that sufficient statistical information is available, namely the wind spectra, integral length scale, and turbulence intensity. For two (and more)-point statistics of the wind turbulence, it is important to account for the correlation between wind vortices especially when simulating the structural response. This can be achieved using the co-coherence function or other approaches.

*Ice Hazard*: Ice loading can have significant effects especially on transmission lines, wind turbines and cable-stayed bridges. Therefore, it should be characterized thoroughly during an ice storm along with other hazards (e.g., wind). The atmospheric icing accretion can be categorized into three main groups, namely precipitation icing, in-cloud icing and hoarfrost (Fikke et al., 2008). Unlike the other categories, the first category (i.e., precipitation icing) includes additionally several ice forms, namely glaze, rime, wet snow and dry snow. Although a circular shape is usually adopted for the ice accretion over specific structures (e.g., overhead lines), the actual ice accretion shape depends on several factors such as the incoming wind speed and direction, the temperature variation and the ice event duration, among others. To statistically characterize the ice accretion, two main approaches can be followed. In the first one, the ice accretion is measured based on specialized equipment (e.g., Ice Rate Meter and on-site cameras). This field-measurement data is then statistically processed to derive the corresponding PDF. In the second approach, several important environmental parameters are first retrieved and statistically represented. Then, ice accretion models are employed to mathematically generate the ice accretion values (usually by assuming a cylindrical shape) (Abdelaal et al., 2019). Several models have been developed in this regard such as the Jones model (Jones et al., 2004) and Makkonen model (Makkonen, 1984). In general, the ice accretion models can be categorized as empirical, numerical, and data-driven models. The latter have recently gained increasing attention due to their high efficiency and accuracy.

### 4.2 Structural characterization

The structural characterization step gives a probabilistic estimation of the *SP* component which is required to evaluate the probability density function, $f(SP)$, in the PBIE framework. Specifically, several structural characteristics need to be represented, including the mechanical and geometrical properties of the structural system which will affect eventually the interaction parameter and structural response. These sets of parameters are usually related to the stiffness, material strength and properties, structural damping and structural shape and geometric configuration which will govern the wind and ice induced loads. Other parameters might also be important depending on the type of structure. For example, in the case of overhead lines, the ability of the conductor to rotate is a key parameter which will govern the total ice deposit. For the case of floating offshore wind turbines which are subjected to several loads such as the wind, waves, currents, and ice, the mooring lines properties have to be included since they affect the structural response (Tuhkuri and Polojärvi, 2018). Additional parameters have also been reported in the literature such as the connectivity, robustness and redundancy which might be necessary to accurately determine the environmental-induced loads and the total response (Barbato et al., 2013). The mechanical properties of passive, active, semi-active or hybrid control technologies should also be included since they modify the overall structural characteristics (e.g., detuning pendulums, tuned mass dampers, and tuned liquid column dampers) (Leroux et al., 2023). It should be noted that the proposed PBIE allows for the consideration of the damage effects on the structural parameters (Fig. 1). In this case, an adaptive $f(SP)$ is required to reflect the changes in the structural parameters.

### 4.3 Interaction analysis

The interaction analysis provides the probabilistic description of the interaction parameter *IP*. It refers essentially to the aerodynamic and aeroelastic phenomena which govern the hazard-induced loads (e.g., ice-wind-induced vibrations for stay cables and overhead lines). Therefore, it characterizes the interaction between the environment and structure (Fig. 3). For instance, in the case of a cable-stayed bridge, the ice-wind-induced cable vibrations are essentially due to: 1. the buffeting loads which are related to the self-induced turbulence signature or by the turbulence of the incoming wind (modified by the momentum exchange between wind and ice particles); 2. the vortex-induced vibration which is caused by the alternate shedding of vortices from two sides of the cables (which might be amplified due to the ice accretion); 3. the galloping mechanism which can lead to large amplitude and low frequency oscillations due to the combined wind and ice effects; and 4. the wake galloping which is caused by variations in drag and across-wind forces for cables in the wake of other structural components. Self-excited forces which are induced by aeroelastic interaction between the wind and the structural motion of specific structures could also be important. Several interaction parameters can be then extracted from the above-mentioned phenomena such as the force coefficients, aerodynamic admittance, flutter derivatives, etc., which are affected by the joint effects of wind and ice (and possibly other hazards) from one side and the structure itself from the other side. Therefore, the *IP* are dependent on the intensity measures *IM* and the structural parameters *SP*. Hence, both *IM* and *SP* uncertainties are required to derive the probability distribution of *IP* (i.e., $f(IP|IM, SP)$).

### 4.4 Structural analysis

In this step, the engineering demand parameter (*EDP*) is probabilistically characterized which represents a specific response of the structural system/element (or non-structural element). For instance, the *EDP* can represent the forces, moments, deflections, stresses, velocities, accelerations, etc. Finite element modeling is usually carried out to retrieve the derived EDPs (Desai et al., 1995). Despite their limitations, experimental (e.g., wind tunnels) and field-measurement approaches, can also provide useful information related to certain EDPs. Recently, data-driven techniques have gained increasing popularity and were successfully applied for the prediction of the system response of several structures (Wen et al., 2022).

### 4.5 Damage analysis

The probabilistic evaluation of the structural damage within the PBIE framework is reflected through the damage measure vector. Specifically, the probability density function of *DM* conditional to *EDP* ($f(DM|EDP)$) is required (Eq. (1)). The *DM* vector is an indicator which quantifies the potential damage that can be caused to a structure due to the involved hazards, and assists in the decision-making step to design and evaluate the structural performance under the considered loads. The selection of the *DM* vector is usually dependent on the structural type and desired performance, hence, selecting a suitable formulation of the *DM* with respect to the *EDP* is crucial since it enables the assessment of the damage conditions associated with the specific values of the engineering demand parameters. Damage states (also denoted as limit states) are usually selected for the *DM* vector to represent the degree of damage that a structure or structural element experiences until it becomes non-functional or completely collapses under various levels of hazard intensities. The damage analysis has been conventionally described in terms of the fragility curves in several applications (Rezaei, 2017). Other studies have also selected the engineering demand parameter as a damage measure (i.e., $EDP = DM$) while representing the system performances by the limit states. The latter can be expressed as the difference between the values of *EDP* and the engineering demand capacity parameters (*ECP*). If the *EDP* is higher than the *ECP*, then the loss of functionality or the collapse can be assumed.

## 4.6 Loss analysis

In the loss assessment step, the conditional complementary cumulative distribution function $G(DV|DM)$ is evaluated. This module usually evaluates the potential economic, social and environmental losses associated with the damage or collapse of a structure (or structural element) due to ice events. Both direct and indirect costs of ice (or related hazards)-induced damage can be estimated in the loss analysis step, including repair or replacement costs, business interruption losses, injuries or loss of life, environmental impact, and loss of community service, etc. Although the loss analysis is usually quantified in terms of monetary values, several loss aspects related to various metrics (e.g., social, environmental or psychological aspects) cannot be easily evaluated. The results of the loss analysis step are usually used in the decision-making process to inform the selection of appropriate design measures, including ice/wind load-resistant materials, structural shape (to mitigate the wind/ice induced loads), and protective measures with a goal of selecting the most cost-effective design approach.

## 4.7 Regional risk analysis

The regional large-scale risk analysis in the PBIE framework involves the evaluation of the ice-related hazards and their induced risk for an entire region rather than focusing on an individual structure. Two possible applications of the regional scale-based PBIE are proposed here. The first one assists in the design of new structures or the retrofitting of existing ones by comprehensively characterizing the ice-related hazards and risks based on the retrieved data. This allows the engineers to identify vulnerable areas and high-risk regions (e.g., galloping favorable conditions) to either avoid them or develop suitable mitigation strategies. In addition, the effects of climate change scenarios can be evaluated on the regional scale by integrating the potential changes in ice events, their induced loads and climate-dependent structural response to ensure the structural resiliency over the long term. The second application of the regional scale-based PBIE involves the real-time risk analysis of the ice events in the studied region to rapidly identify the high-risk areas and the vulnerable structures. This aspect has several implementations including those related to the transportation networks, utility systems and emergency response services.

Considering the high computational costs related to each module in the PBIE, it is not feasible to use physics-based models to estimate the risk. In addition, due to the inherent uncertainties within each PBIE module which involves the application of Monte Carlo techniques (or their variants) to estimate the risk, the numerical computational cost will further increase. Therefore, it is important to rely on alternative approaches which will assist in the determination of the vulnerable regions. Here, data-driven techniques can be integrated within the PBIE framework. To illustrate this concept, the case of the electrical grid subjected to ice storm yielding two hazards (i.e., wind and ice for simplification) will be presented here as illustrated in Fig. 4.

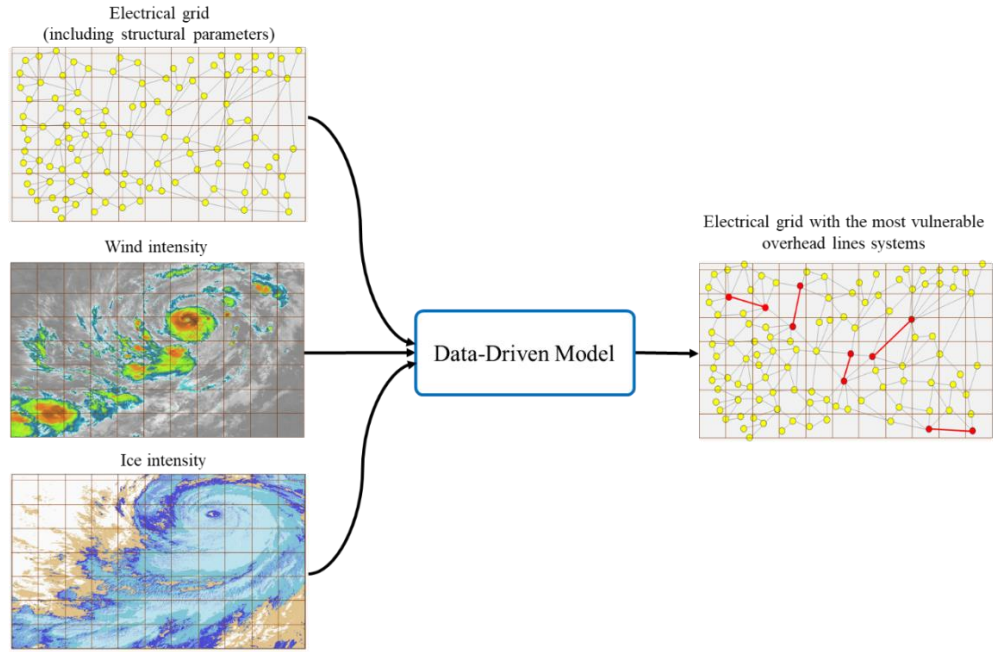

**Fig. 4**. Real-time risk assessment using the regional scale-based PBIE.

As indicated in Fig. 4, the electrical topology should be known along with an approximate estimation of the structural properties of the tower-line overhead system. The hazard maps are also required by the framework. With these data, taken as inputs to an already trained data-driven technique (e.g., machine learning) the risk values can be quickly generated based on the model which will identify the most vulnerable elements in the grid. As the hazard intensities evolve with time, the risk values will also change accordingly. Once the high-risk areas have been determined, then advanced and accurate models (e.g., finite elements) can be used to accurately simulate the risk for more precise analysis. It should be noted that the trained data-driven model can be based on simplified representations of each PBIE module for a convenient training process. However, in this case, the generated risk will be highly approximated and should be complemented with more information to update its simulation by using several techniques such as data assimilation.

## 5. Case Study

In this section, the proposed PBIE framework will be implemented to evaluate the vertical galloping-induced risk on iced conductors. Specifically, the PBIE will be developed for two scales, namely the structural (local) scale to simulate a design scenario using past environmental data and the regional (large) scale to predict the risks associated with an actual incoming ice storm in real-time. To characterize the dynamic response of a given transmission line system subjected to both wind and ice loads, a simplified semi-empirical approach is employed to estimate the galloping amplitude of a transmission line system (Lu and Chan, 2009; Lu and Chakrabarti, 2023). More advanced theories can be readily integrated within the proposed framework. However, since the main objective of this case study is to demonstrate the capabilities of the proposed methodology, which leverages the machine learning properties, it was deemed sufficient to employ the linear theory of a suspended cable. In addition, the galloping-induced risk will be simply expressed in terms of the double galloping amplitude $Y$. A brief description of the galloping instability, along with the galloping conditions and dynamic model, will be presented in the following sections.

## 5.1 Galloping theory of the transmission line system

### 5.1.1 Galloping instability and conditions

Various types of unsteady wind-induced loading and associated oscillations affect the transmission lines (Lou et al., 2023). With the presence of ice, the wind-induced oscillations might significantly increase. Although Aeolian vibrations and wake-induced oscillations might affect the dynamic performance of transmission lines, especially through the fatigue mechanism, they are usually characterized by small amplitude oscillations. Galloping, on the other hand, can lead to large amplitude and low frequency oscillations due to the combined wind and ice effects (McComber and Paradis, 1998).

To determine the onset of the galloping instability, two main criteria need to be fulfilled (Den Hartog, 1932). The first criterion is based on the Den Hartog coefficient, $a_g$, which has to be negative:

$$a_g = C_d + \frac{dC_l}{d\alpha} < 0 \tag{4}$$

where $C_d$ = aerodynamic drag coefficient; $C_l$ = aerodynamic lift coefficient; and $\alpha$ = Angle of Attack (AoA). The second criterion is governed by the total system damping, $\xi$, such that:

$$\xi = \xi_s + \xi_a < 0 \tag{5}$$

where $\xi_s$ = structural damping; and $\xi_a$ = aerodynamic damping. To achieve the second condition, the wind speed has to be higher than a critical wind speed value, also denoted as $V_c$ and expressed as:

$$V_c = -\frac{4.\omega_n.\xi_s.m_s}{\rho_0.d_c.a_g} \tag{6}$$

where $\omega_n$ = natural angular frequency; $m_s$ = mass of the conductor; $\rho_0$ = density of ice; and $d_c$ = Conductor's Diameter.

### 5.1.2 Dynamic model of the transmission line system

In this study, a simplified semi-empirical equation is employed to estimate the galloping amplitude $A_G$ of a transmission line system (Lu and Chan, 2009; Lu and Chakrabarti, 2023). The galloping equation is formulated based on extensive parametric studies conducted using a hybrid finite element and three-degree-of-freedom (3DOF) computer software. This design equation is intended to predict $A_G$ for both single and bundle conductors with either suspension or dead-end supports, making it suitable for real-world engineering applications. Calibration of the design equation was achieved using field galloping data observed in North America. The results obtained from the proposed formula demonstrate good agreement with those generated by the numerical approach. Building upon this simplified approach, the galloping amplitude in plunge can be expressed by the following equation:

$$A_G = G/f \tag{7}$$

where $G$ = galloping factor; and $f$ = conductor's natural frequency in plunge during galloping. Analysis of field data on galloping observed in North America suggests that a value of $G = 0.75 m/s$ can be adopted as a design constant for North American applications, particularly in the absence of site-specific galloping data. This selection of $G$ allows the proposed design equation to accurately encompass the available field measurements of galloping in North America. The natural frequency of a sagged conductor span in plunge with both ends fixed can be determined as:

$$f = \frac{k}{L}\sqrt{\frac{H}{m}} \tag{8}$$

where $L$ = span length of the conductor; $m$ = mass per unit length of the conductor with ice accumulation; $H$ = static conductor tension; and $k$ = a parameter determined by the number of galloping loops. In the case of two-loop galloping, the parameter $k$ takes a value of 1. However, for one-loop galloping, $k$ needs to be determined by solving the following equation (Irvine and Caughey, 1974):

$$\tan(k\pi) = k\pi - \frac{4(k\pi)^3}{\lambda^2} \tag{9}$$

where the parameter $\lambda$ is defined by the following equation:

$$\lambda = \frac{mgL}{T}\sqrt{K_c \frac{L}{T}} \tag{10}$$

where $K_c$ = conductor stiffness which is expressed for a conductor span with both ends fixed as:

$$K_c = \frac{EA}{L} \tag{11}$$

where $E$ = equivalent elastic modulus of the conductor; and $A$ = cross sectional area of the conductor. The double galloping amplitude $Y$ can be directly determined from the single galloping amplitude $A_G$ by $Y = 2A_G$.

### 5.2 Application

#### 5.2.1 Local scale application

In this section, the PBIE framework will be applied to evaluate the vertical galloping-induced risk on an iced conductor to simulate a design scenario using past environmental data. Since a simplified semi-empirical formula is selected here to evaluate the dynamic response of the transmission line system, no damage analysis will be carried out. Instead, the limit states are defined in terms of the double galloping amplitude $Y$. Specifically, a limit state function ($g_l$) is introduced here for assessing the risk of failure. This function compares the engineering demand parameter obtained from the structural analysis with the limit state capacity (resistance) of the structure for each potential damage state. Mathematically, the limit state function is defined as $g_l$ = Demand - Capacity. A positive value of $g_l$ ($g_l > 0$) indicates that the demand exceeds the capacity, signifying a failure condition. Conversely, a non-positive value ($g_l \leq 0$) suggests the structure remains within its safe zone. Three limit states have been selected to quantify the risk categories as indicated in Table 1.

**Table 1**. Summary of the risk levels.

| Risk level | Risk definition | Limit state capacity |
|---|---|---|
| 1 | Low | $2\ m$ |
| 2 | Moderate | $3\ m$ |
| 3 | High | $4\ m$ |

*5.2.1.1 Hazard analysis & structural characterization*

This study utilizes environmental data, including wind and ice accretion, collected from Hydro-Québec's comprehensive ice measurement network (Laflamme, 2004). Operational since fall 1974, this network utilizes unique '*glacimètre*' stations deployed at 54 locations (half within the St. Lawrence Valley). These autonomous stations collect detailed ice accretion data on eight cylindrically shaped surfaces facing each cardinal direction. The *glacimètre*'s design allows for characterization of icing events by type, accumulation on specific orientations, thickness, and start/end times. This rich data serves as a critical source for developing robust design criteria for overhead transmission line systems, making Hydro-Québec's network one of the most comprehensive ice accretion databases in North America. For the local scale application, the mean wind speed and equivalent ice accretion ($R_{eq}$) data for the period 1975-2020 were acquired from four stations: Lachute (45.65°N, -74.33°W), Montreal A (45.46°N, -73.75°W), Fortierville (46.46°N, -72.01°W), and Honfleur (46.68°N, -70.86°W). To analyze the wind and ice distributions for the period 1975-2020, a generalized extreme value (GEV) distribution was employed. In addition, a copula approach is adopted in this study to account for the potential correlation between wind and ice and their combined effects on overhead transmission lines.

It should be noted that a crescent shape of the accreted ice is usually observed for fixed conductors. Therefore, it is important to realistically represent the ice profile on the conductor since it affects the wind-induced aerodynamic loads. Similar to previous studies (Rossi et al., 2020, Davalos et al., 2023), a semi-elliptical shape for the accreted ice is taken here to model the ice deposit on the conductor. Hence, the equivalent radial ice accretion is transformed to an ice deposit '*t*' using the following approximated formula:

$$t = 4.R_{eq}.\left(1 + \frac{R_{eq}}{d_c}\right) \qquad (12)$$

To consider the relative dimension of the ice deposit (*t*) with respect to the conductor diameter ($d_c$), the ice eccentricity (*e*) will be integrated in the proposed framework which is defined as:

$$e = {t}/{d_c} \qquad (13)$$

To evaluate the galloping-induced risk through the proposed probabilistic framework, it is important to consider the uncertainties related to the structural parameters. Since a simplified dynamic model has been utilized for the conductor, only a few parameters will be considered as random variables while the others are assumed deterministic. Specifically, the conductor diameter is assumed as normally distributed with a mean value of 23.55 mm and a coefficient of variation of 0.05 (Truong and Kim, 2017). In addition, the modulus of elasticity is assumed to follow a log-normal distribution with a mean value of 70 GPa and a coefficient of variation of 0.03 (Fu et al., 2020).

*5.2.1.2 Interaction analysis*

The interaction analysis is captured in this study through the force coefficients (e.g., drag coefficient). It should be noted that the analytical/numerical determination of wind forces on a bluff body (e.g., crescent shape of the accreted ice) is quite difficult. Geometrically scaled models are often used in practice to obtain the pressure (or force) coefficients through wind tunnel tests (Rossi et al., 2020). In particular, the force coefficients which can be retrieved from experimental tests, are nondimensional quantities which are dependent on a number of variables related to the geometry of the body and the upwind flow characteristics. Computational Fluid Dynamics (CFD) modeling can also be applied to retrieve the force coefficients (Li et al., 2016), however, they are usually computationally expensive. To overcome some of the aforementioned issues, an artificial neural network (ANN) is trained here based on experimental wind tunnel tests carried out by (Rossi et al., 2020). The ANN model predicts the aerodynamic drag and lift coefficients, which are required to evaluate the Den Hartog coefficient $a_g$, based on three inputs, namely the angle of attack ($\alpha$) of the free stream velocity with respect to the cable's section, the ice eccentricity ($e$), and the incoming wind speed ($U$). The ANN model employs a two-layer hidden architecture with 20 neurons in each layer. The Xavier initialization technique is used for the weight initialization and a Relu function is selected for the activation function. In addition, the bias vector is initialized to zero. The Adam optimizer with a value of 0.001 for the learning rate is employed for the ANN model. A schematic of the ANN Architecture is depicted in Fig. 5.

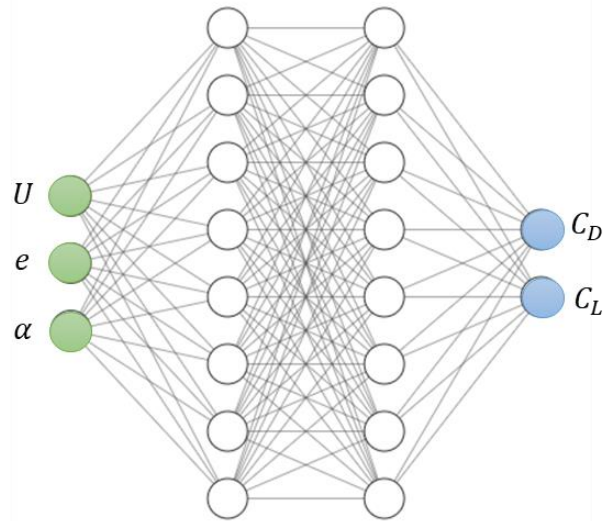

**Fig. 5**. ANN architecture for the prediction of the aerodynamic drag coefficient.

The performance of the proposed ANN model is assessed through the mean squared error (MSE) and R-squared parameter ($R^2$). The obtained MSE ($R^2$) values for the training [85% of the data] and testing [15% of the data] sets are 0.02 (95%) and 0.03 (92%), respectively which indicate a good performance of the trained model.

*5.2.1.3 Structural analysis*

The structural analysis is based on the simplified semi-empirical approach of Sect. 5.1.2. The necessary parameters needed for the analysis are: $L = 172.4\ m$, $E = 70\ GPa$, $sag = 3.39\ m$, $m_s$=1.3 kg/m (mass of the conductor), and $A_c = 4.35 * 10^{-4}\ m^2$ (area of the bare conductor). If the two main criteria for the galloping onset have been fulfilled (i.e., Eqs. 4 and 5), then the double galloping amplitude $Y$ can be calculated.

*5.2.1.4 Risk analysis*

Based on the selected three limit states (Table 1) which identify the risk categories, the galloping-induced risk has been calculated using the Monte Carlo technique (10,000 samples). The obtained results at the four selected stations are summarized in Fig. 6.

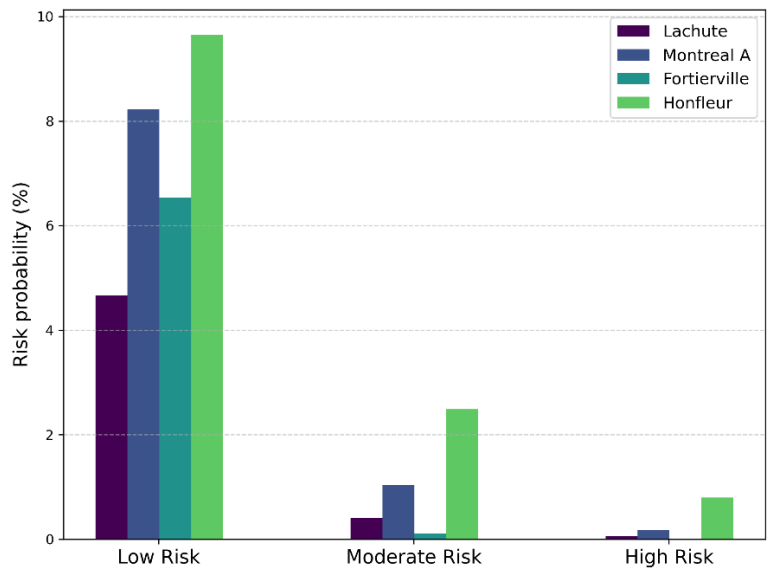

**Fig. 6** Results of the galloping-induced risk analysis at the four selected stations

Figure 6 illustrates that the probabilities of failure due to galloping excitation are non-negligible for the chosen limit states at all four stations. As anticipated, the low-risk category exhibits higher probabilities compared to moderate and high-risk categories due to the lower limit state capacity (2 m) for low risk. Notably, Honfleur station appears most susceptible to galloping, with significantly higher probabilities across all risk categories (low: 9.65%, moderate: 2.49%, extreme: 0.8%). Conversely, Lachute exhibits the lowest risk for the low limit state (4.66%). However, for moderate and high-risk categories, Fortierville presents the lowest probabilities (0.11% and 0.003%, respectively). It is important to acknowledge that these probabilities are approximations, and more sophisticated aerodynamic/dynamic models could provide more accurate results. Additionally, the risk probabilities shown in Fig. 6 were generated by assuming the angle of attack to be a random variable with a uniform distribution, as this information is missing in the database. If a value of $\alpha = 180°$ is used, which is more likely to cause galloping (other values can also be used if this information is known), then the resulting risk probabilities will increase significantly. For example, for the Honfleur station, the estimated probabilities for low, moderate, and high risk are 63.7%, 8.17%, and 2.64%, respectively, when $\alpha = 180°$.

## 5.2.2 Regional scale application

In this section, the PBIE framework will be applied to estimate the vertical galloping-induced risk on iced conductors over large geographical areas. Specifically, it will explore how the framework can be applied in real-time to predict the risk posed by an approaching ice storm. Similar to the local scale, the simplified semi-empirical approach of Sect. 5.1.2 will be employed to calculate the double galloping amplitude $Y$, employing similar limit states as shown in Table 1. To illustrate the effectiveness of the PBIE framework in simulating the real-time risk over an extended area, two real ice storms which occurred in the Quebec province will be selected. The necessary environmental data for the proposed framework corresponding to the selected storms will be retrieved from the Hydro-Quebec database (Périard, 2018). The first ice storm occurred between February 4th and 5th, 2019, impacting Quebec's Outaouais, St. Lawrence Valley, Lower St. Lawrence, and Baie-des-Chaleurs regions. The Montmagny area received the most ice accumulation, with up to 14mm. The ice deposits were short-lived, lasting 1 day or less at most stations. Winds were calm during the event, and the maximum amount of rain, at 11.9mm, was recorded at the Lachute station in Outaouais. Temperatures ranged between -2.0°C and -8.0°C. The storm caused some power outages, road accidents, and school closures, although transportation and distribution networks remained largely unaffected. The second ice storm event, which occurred between April 8th and 9th, 2019, impacted Laval, Lanaudière, and Laurentides regions in Quebec. Most reporting stations recorded an average equivalent ice thickness of less than 5 mm. There were no reported transportation disruptions. Temperatures ranged from -1.0°C to -2.0°C with northeast winds of 10 to 45 km/h. Ice deposits persisted for less than 24 hours to 3 days. The storm caused widespread power outages, leaving over 475,000 Hydro-Québec customers without electricity for several days. The measured equivalent radial ice accretion values for the two storms are depicted in Fig. 7.

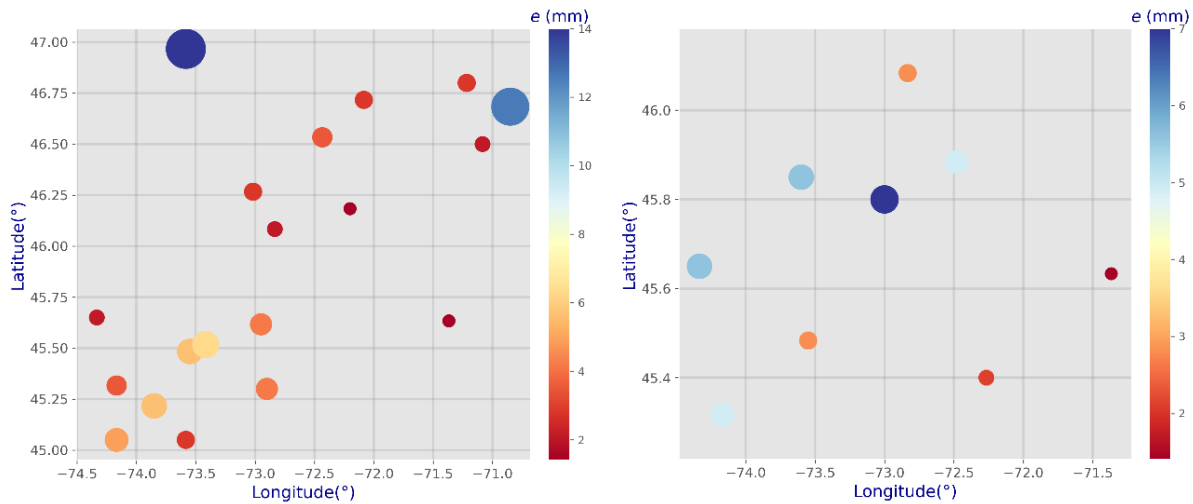

**Fig. 7** Equivalent radial ice thickness values over the selected stations due to storm 1 (left) and storm 2 (right)

The kriging approach is then utilized to generate the wind and ice maps which are required for the risk assessment.

To generate the risk maps due to the ice storm event, the Monte Carlo technique can be repeated for each grid point of the extended area (assuming that the transmission line systems are located within each grid). However, proceeding this way ('naive' way) will require substantial computational time which is not convenient for real-time risk assessment. To alleviate the high-computational cost of the Monte Carlo technique, an ANN model will be developed to efficiently predict the galloping-induced risk. The neural network has two inputs, namely the mean wind speed (*U*) and the ice eccentricity (*e*). In addition, the network has three outputs which represent three risk levels [1 (low); 2 (moderate); 3 (high)] as defined in Table 1. The network architecture utilizes a three-layer hidden structure, with each layer containing 40 neurons. A Relu activation function has been selected for the hidden layers and a sigmoid function was chosen for the output layer. Adam optimizer and a learning rate of 0.01 have been selected for the ANN training. The architecture of the ANN is depicted in Fig. 8.

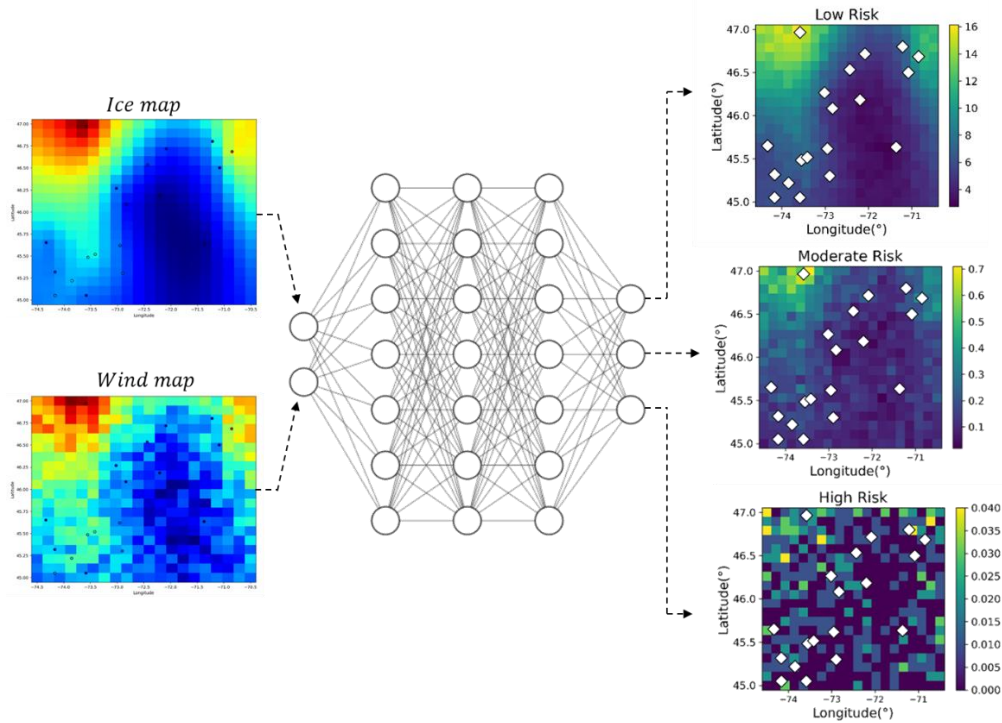

**Fig. 8**. ANN architecture for the prediction of the galloping-induced risk

The necessary data for the model training have been generated using the Monte Carlo simulation. Specifically, a total of 500 combinations of the input parameters (*u, e*) have been generated and their corresponding risk values have been calculated using the Monte Carlo technique. The obtained MSE ($R^2$) values for training (85% of the generated data) and testing (15% of the generated data) sets are 0.006 (91%) and 0.01 (82%), respectively. These obtained results indicate that the model is well trained and can be applied to predict the galloping-induced risk levels. It should be noted the angle of attack (*α*) was treated as an unknown variable with a uniform probability distribution. However, if the actual angle of attack is known, it can be directly incorporated as an input into the ANN model. An example will be provided to demonstrate this concept. Based on the trained ANN model, three risk maps (representing the three selected risk levels) have been generated using the environmental parameters corresponding to the selected ice storm.

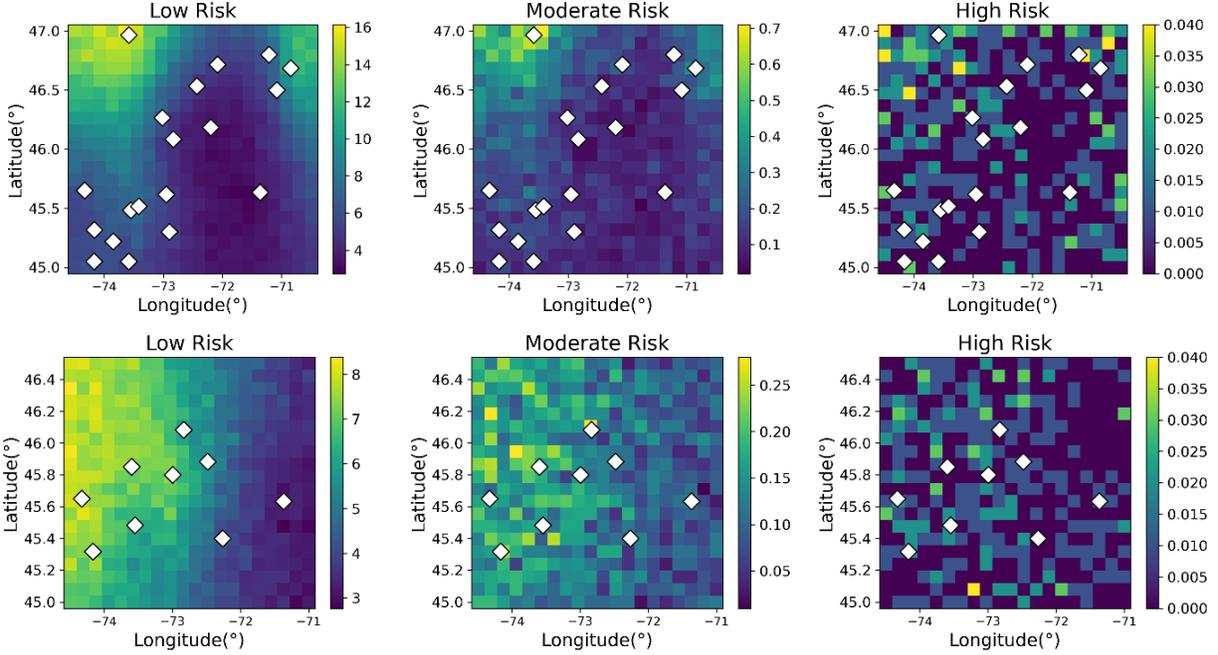

**Fig. 9**. ANN-based galloping-induced risk maps for Storm #1 (top) and Storm #2 (bottom) [Diamond symbols indicate station locations]

The results indicate areas of potential high galloping risk highlighted in yellow. For the first storm, these vulnerable areas are concentrated around 47° N latitude and between -73° W and -75° W longitude. Similarly, for the second storm, high susceptibility to galloping is found between -73° W and -75° W longitude and latitudes ranging from 45.2° N to 46.4° N. These risk maps allow the user to prioritize areas for further study. By focusing on the high-risk zones (shown in yellow) with advanced numerical models, the galloping risk can be predicted efficiently. This targeted approach saves significant time compared to analyzing the entire region, which would be computationally expensive. Therefore, several mitigation actions can be taken especially within those affected areas to alleviate the negative effects of galloping which can jeopardize the structural integrity of overhead lines. It should be noted that several simplifying assumptions have been taken here to generate the risk maps, however, with more accurate dynamic models and environmental data, it is expected that the prediction accuracy will improve significantly.

As previously mentioned, if the angle of attack is known, it can be used as an input for the ANN model. For example, if $\alpha = 180°$, which is known to yield a negative Den Hartog coefficient (Rossi et al., 2020) and may lead to galloping, new risk maps can be generated, as shown in Fig. 10.

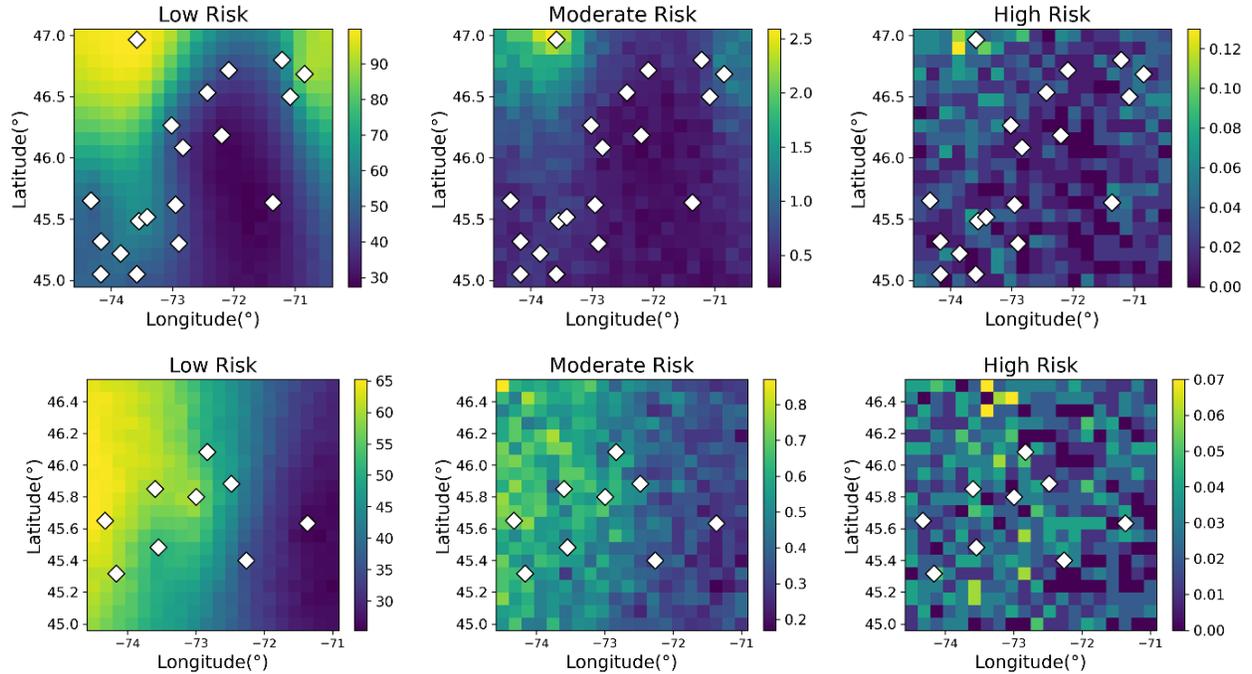

**Fig. 10**. ANN-based galloping-induced risk maps with $\alpha = 180°$ for Storm #1 (top) and Storm #2 (bottom)

As it can be concluded from Fig. 10, higher risk values for the three categories and the two storms have been generated, as expected, since the selected angle of attack yields a negative Den Hartog coefficient and hence a higher likelihood of leading to galloping. To highlight the efficiency of the proposed ANN-based framework, the same risk maps are generated using directly the Monte Carlo simulation 'naive simulation' for each grid point. The results indicate that while the ANN-based framework almost instantly generate the risk maps (with an approximate runtime of 0.007 seconds), the naive way requires up to 3,000 seconds to generate the same maps on a regular personal computer (CPU @ 3.20 GHz). Therefore, the proposed model can help in identifying efficiently the vulnerable structures in high-risk areas and can be readily integrated within real-time risk assessment platforms.

The proposed framework offers broad applicability to structures such as wind turbines, bridges, and overhead transmission lines. While the current case study focuses on a simplified overhead transmission line system, it demonstrates the core functionality, particularly the integration of data-driven techniques for real-time risk assessment. However, some limitations need to be addressed to fully explore the functionality of the proposed framework. First, while the simplified case study can predict the galloping amplitude of power lines due to ice storms, it doesn't take into account the potential damage/loss this might cause. This is because including complex damage/loss calculations would require very demanding computer models. As a result, the framework's current output might be useful for engineers but less informative for decision-makers who need a clearer picture of the potential risks. To address this, incorporating modules to estimate damage or loss would make the framework more valuable for a wider range of users. However, the framework, as illustrated in Fig. 1, can be easily extended to incorporate damage and loss assessment when appropriate models are available. Secondly, the case study uses a simplified, uniform grid network for illustration, but real-world applications involve specific electrical layouts with varying characteristics. To generate more reliable risk assessments, incorporating these network details would be beneficial. The framework's applicability could be further explored by applying it to different structures, such as wind turbines and bridges. For large-scale risk assessments involving numerous structures, exploring advanced data-driven techniques like dimensionality reduction (e.g. Saviz Naeini and Snaiki, 2024), knowledge-enhanced deep learning (e.g., Snaiki and Wu, 2019, 2022), and image processing could significantly enhance the simulations.

## 6. Concluding Remarks

In this study, a novel performance-based ice engineering (PBIE) framework has been developed. The proposed methodology is adaptable for different scales since it is capable of ensuring effective design and rehabilitation of new and existing structures (local scale-based PBIE), respectively and is also suitable for the rapid evaluation of ice storms-induced risk over extended geographical areas (regional scale-based PBIE). The risk estimation in the local scale-based PBIE relies on the total probability theorem in which the risk is disaggregated into: 1. hazard analysis; 2. structural characterization; 3. interaction analysis; 4. structural analysis; 5. damage analysis; 6. loss analysis; and 7. decision-making. The local scale-based PBIE can also accommodate the multi-hazard nature of ice storms and time-varying factors (e.g., climate change and aging). On the other hand, the regional scale-based PBIE leverages the superior capabilities of surrogate models (e.g., machine learning) to quickly predict the ice storm-induced risk over a large area using information about the hazard intensities and an approximation of the structural parameters along with the network topology. The proposed PBIE (both local and regional scales) was illustrated through a simplified application consisting of the risk assessment of galloping-induced vibrations on iced conductors subjected to both wind and ice loads. The results showed that the interaction between the wind speed and ice accretion can significantly affect the onset of the galloping instability which will also alter the iced conductor's dynamic response and eventually the induced risk. On the other hand, by leveraging machine learning techniques (here an artificial neural network was used) the risk maps due to the event of an ice storm can be almost instantly generated compared to the standard way, based on Monte Carlo simulations, which requires substantial computational time and hence cannot be employed for real-time risk applications.

**Funding:** This work was supported by the Natural Sciences and Engineering Research Council of Canada (NSERC) [grant number CRSNG RGPIN 2022-03492].

**Conflict of interest:** The author declare that he has no conflict of interest.